\documentclass{ws-procs975x65}            
\begin{document}
\title{Nonlinear gravitational waves as dark energy in warped spacetimes}
\author{Reinoud J. Slagter$^*$}
\address{ASFYON and Department of Physics, University of Amsterdam,\\
Bussum, The Netherlands \\
$^*$E-mail: info@asfyon.com}
\begin{abstract}
On a warped five-dimensional Friedmann-Lema\^{\i}tre-Robertson-Walker(FLRW) spacetime, dark energy can be induced by a U(1) scalar-gauge field on the brane. We consider  a zero effective cosmological constant, i.e., the Randall-Sundrum(RS) fine-tuning and no bulk matter fields. The standard model fields interact via the bulk Weyl tensor and cause brane fluctuations. Due to the warp factor, disturbances don't fade away during the expansion of the universe. The late-time behavior could be significant deviate from the standard evolution of the universe. The effect is triggered by the time-dependent part of the  warp factor. The self-gravitating cosmic string builds up a huge mass per unit length in the bulk and can induce massive KK-modes felt on the brane. From a nonlinear perturbation analysis it is found that the effective Einstein equations contain a "back-reaction" term on the righthand side  caused by the projected 5D Weyl tensor and can act as a dark energy term. The propagation equations to first order for the  metric components and scalar-gauge fields show explicit $\varphi$-dependency.
\end{abstract}
\bodymatter
\section{Introduction}\label{aba:sec1}
Recent observations provide strong evidence for the acceleration of our universe. The explanation of this remarkable phenomenon is rather difficult: one needs a dark energy field, which could be the cosmological constant.
However, there is a huge discrepancy between the observed value of the cosmological constant and the value obtained from the vacuum energy density. Moreover, there is an incredibly fine-tuning  between the energy density of this dark energy and the energy density of matter.
It is a great challenge to find other explanations for the late-time acceleration, without the need for a cosmological constant. Physicists speculate that extra spatial dimensions could exist in addition to our ordinary 4-dimensional spacetime.
These theories can be used to explain several of the shortcomings of the Standard Model, i.e., the unknown origin of dark energy and the weakness of gravity (hierarchy problem). Recently there is growing interest in the so-called brane world models\cite{IA:1998,RS:1999}. In these models, the weakness of gravity might be fundamental.
Compact objects, such as black holes and cosmic strings, could have tremendous mass in the bulk, while their warped manifestations in the brane can be consistent with observations. The energy scale of the cosmic string, $G\mu$, could be warped down to GUT scale, even if its value was at the Planck scale.
Cosmological cosmic strings can also be investigated on a FLRW background. However, the string-cosmology spacetime essentially looks like a scaled version of a string in a vacuum spacetime.\cite{Greg:1989}
Static and time-dependent solutions of the U(1) gauge string on a warped FLRW spacetime show significant deviation from this classical solution in 4D\cite{Slag2:2014,Slag3:2015}.
While the inflaton field plays a crucial role in the early stage of our universe, it could play a comparable role at much later times, if we modify gravity by considering warped brane-world models. It could be possible that there exists a correlation between the accelerating universe and large extra dimensions in brane-world models.
Wavelike disturbances triggered by the huge mass of the cosmic string in the bulk, could have observational effects on the brane. One conjectures that these disturbances could act as an effective dark energy field.
In this paper we examine the possibility that cosmological cosmic strings, in contrast with the 4-dimensional counterpart model,\cite{Greg:1989} could contribute to a dark radiation and triggers the acceleration or deceleration of the universe. This conjecture is founded by a multiple-scale analysis.
\section{The 5-dimensional Model}
We consider here the 5-dimensional Einstein equations with only a bulk cosmological constant $\Lambda_5$\cite{shir:2000}
\begin{equation}
{^{5}\!G}_{\mu\nu}=-\Lambda_5{^{5}\!g}_{\mu\nu}+\kappa_5^2 \delta(y)\Bigl(-\Lambda_4 {^{4}\!g}_{\mu\nu}+{^{4}\!T}_{\mu\nu}\Bigr),\label{eqn1}
\end{equation}
with $\kappa_5= 8\pi {^{5}\!G}= 8\pi/{^{5}\!M}_{pl}^3$, $\Lambda_4$ the brane tension, ${^{4}\!g}_{\mu\nu}={^{5}\!g}_{\mu\nu}-n_\mu n_\nu$,
and $n^\mu$ the unit normal to the brane. We consider here the matter field ${^{4}\!T}_{\mu\nu}$ confined to the brane, i.e., the U(1) scalar-gauge field, written in the form \cite{Garf:1985} $\Phi=\eta X(t, r)e^{i\varphi}, A_\mu =\frac{1}{e}(P(t, r)-1)\nabla_\mu\varphi$,
with $\eta$ the vacuum expectation value of the scalar field and potential $V=\frac{1}{8}\beta(X^2-\eta^2)^2$.
Let us consider the cylindrically symmetric warped spacetime
\begin{equation}
ds^2 = {\cal W}(t, r, y)^2\Bigl[e^{2(\gamma(t, r)-\psi(t, r))}(-dt^2+ dr^2)+e^{2\psi(t, r)}dz^2+ r^2 e^{-2\psi(t, r)}d\varphi^2\Bigr]+ dy^2,\label{eqn2}
\end{equation}
with ${\cal W}$ the warp factor and $y$ the bulk space coordinate.
One can analyse the behavior of gravitational waves generated by the local strings on the expansion of the universe. One usually considers the zero-thickness limit of non-singular spacetimes containing a cylindrical distribution of stress-energy embedded on a cosmological background. In general, however, the late-time behavior of non-vacuum spacetimes, show conical behavior, not desirable. It was found\cite{Greg:1989} that a U(1) cosmic string can be embedded into a flat FLRW spacetime along the polar axis.
It turns out that when the width of the strings is smaller than the Hubble radius, the disturbances are negligible.
On a warped 5-dimensional spacetime, they could survive and the effective brane spacetime is non-conical.\cite{Slag2:2014}
From the 5D equations one obtains the solution for ${\cal W}(t, r, y)$ with two branches:\cite{Slag3:2015}
\begin{equation}
{\cal W}(t, r, y)=\pm\frac{e^{\sqrt{-\frac{\Lambda_5}{6}}(y- y_0)}}{\sqrt{\tau r}}\sqrt{\Bigl(d_1 e^{(\sqrt{2\tau})t}-d_2e^{-(\sqrt{2\tau})t}\Bigr)\Bigl(d_3 e^{(\sqrt{2\tau})r}-d_4e^{-(\sqrt{2\tau})r}\Bigr)}, \label{eqn3}
\end{equation}
with $\tau, d_i$ some constants. In general, ${\cal W}$ can possess a saddle-point or  extremal values.
The modified Einstein field equations induced on the brane  become\cite{shir:2000}
\begin{equation}
{^{4}\!G}_{\mu\nu}=-\Lambda_{eff}{^{4}\!g}_{\mu\nu}+\kappa_4^2 {^{4}\!T}_{\mu\nu}+\kappa_5^4{\cal S}_{\mu\nu}-{\cal E}_{\mu\nu}.\label{eqn4}
\end{equation}
${\cal S}_{\mu\nu}$ is  the quadratic term in the energy-momentum tensor and ${\cal E_{\mu\nu}}$
is a part of the 5D Weyl tensor and carries information of the gravitational field outside the brane.
From the Einstein equations and scalar-gauge field equations one obtains a set of PDE's, which can be solved numerically.\cite{Slag2:2014}
The time dependent part of the warp factor  causes disturbances of the order much larger than the expected values in the 4D case.
Moreover, it was found that for an extremum of $W_1$, the solution diverges. This could have a significant influence on a transition from acceleration to a deceleration or vice versa.\cite{Slag3:2015}
\section{The High-Frequency  Approximation}
We consider fields $V_i$ in point x on a manifold M dependent  on different scales $(x_\mu, \xi, \chi , ...)$:\cite{Choc:1977,Slag4:1986}
\begin{equation}
V_i=\sum_{0}^{\infty} \frac{1}{\omega^n} F_i^{(n)}(x_\mu,\xi,\chi ,...)\label{eqn5},
\end{equation}
with $\omega>>1$, $\xi =\omega \Theta(x_\mu)$, $\chi =\omega \Pi(x_\mu), ...$ and $\Theta, \Pi , ...$  scalar (phase) functions on M. The small parameter $\frac{1}{\omega}$ can be the ratio of the characteristic wavelength of the perturbation to the characteristic dimension of the background (or the ratio of the extra dimension y to the background dimension).
We expand
\begin{eqnarray}
g_{\mu\nu}=\bar g_{\mu\nu}(x_\mu)+ \frac{1}{\omega}h_{\mu\nu}(x_\mu,\xi,\chi)+\frac{1}{\omega^2}k_{\mu\nu}(x_\mu,\xi,\chi) + ... \cr \Phi=\bar\Phi(x_\mu) +\frac{1}{\omega}\Psi(x_\mu, \xi, \chi)+\frac{1}{\omega^2}\Xi(x_\mu, \xi, \chi)+...
\cr  A_\mu=\bar A_\mu (x_\mu)+\frac{1}{\omega}B_\mu (x_\mu,\xi ,\chi) +\frac{1}{\omega^2}C_\mu (x_\mu,\xi,\chi) +...,\label{eqn6}
\end{eqnarray}
with $\bar g_{\mu\nu}$ the background metric and $\bar\Phi, \bar A_\mu$ the background scalar and gauge fields.
Let us now define
\begin{eqnarray}
\frac{d g_{\mu\nu}}{d x^\sigma}=g_{\mu\nu,\sigma}+\omega l_\sigma \dot g_{\mu\nu} \qquad g_{\mu\nu,\sigma}\equiv \frac{\partial g_{\mu\nu}}{\partial x^\sigma}\qquad
 \dot g_{\mu\nu}\equiv \frac{\partial g_{\mu\nu}}{\partial \xi},\label{eqn7}
\end{eqnarray}
with $l_\mu \equiv \frac{\partial \Theta}{\partial x^\mu}$. So we consider, for the time being, only rapid variation in the direction of $l_\mu$  transversal to the sub-manifold $\Theta$ = constant.
We can  expand the several relevant tensors, for example,
$\Gamma_{\mu\nu}^\alpha =\bar \Gamma_{\mu\nu}^\alpha +\Gamma_{\mu\nu}^{\alpha (0)} +\frac{1}{\omega}\Gamma_{\mu\nu}^{\alpha (1)}+...$ and
$R^\sigma_{\mu\tau\nu}=\omega R^{(-1)\sigma}_{\mu\tau\nu}+\bar R^\sigma_{\mu\tau\nu}+R^{(0)\sigma}_{\mu\tau\nu} +.... $,
with
$\Gamma^{\sigma (0)}_{\mu\nu}=\frac{1}{2}\bar g^{\beta\sigma}\bigl( l_\mu \dot h_{\beta\nu}+l_\nu\dot h_{\beta\mu}-l_\beta \dot h_{\mu\nu}\bigr)$ and
$\Gamma^{\sigma (1)}_{\mu\nu}=\frac{1}{2}\Bigl(h^\sigma_{\mu :\nu}+h^\sigma_{\nu :\mu}-h_{\mu\nu}^{: \sigma}\Bigr)-\frac{1}{2}\Bigl(l_\nu \dot k_\mu^\sigma + l_\mu \dot k_\nu^\sigma -l^\sigma \dot k_{\mu\nu}\Bigr)-h_\rho^\sigma \Gamma^{\rho (0)}_{\mu\nu}$.
We substitute  the expansions into the effective brane Einstein equations Eq.~(\ref{eqn4}) and subsequently put equal zero the various powers of $\omega$. We then obtain a system of partial differential equations for the fields $\bar g_{\mu\nu}, h_{\mu\nu}, k_{\mu\nu}$ and the scalar gauge fields $\bar \Phi, \Psi, \Xi, \bar A_\mu, B_\mu \equiv[B_0,B_1,0,B,0]$ and $C_\mu$. The $\omega^{(-1)}$ equation becomes
\begin{equation}
{^{4}\!G_{\mu\nu}^{(-1)}}=-{\cal E}_{\mu\nu}^{(-1)},\label{eqn8}
\end{equation}
and the $\omega^{(0)}$ equation
\begin{equation}
{^{4}\!\bar G_{\mu\nu}}+{^{4}\!G_{\mu\nu}^{(0)}}=-\Lambda_{eff}{^{4}\!\bar g_{\mu\nu}}+\kappa_4^2 \bigl({^{4}\!\bar T_{\mu\nu}}+{^{4}\!T_{\mu\nu}^{(0)}}\bigr)
+\kappa_5^4\bigl(\bar {\cal S}_{\mu\nu}+{\cal S}_{\mu\nu}^{(0)}\bigr)-\bigl(\bar{\cal E}_{\mu\nu}+{\cal E}_{\mu\nu}^{(0)}\bigr).\label{eqn9}
\end{equation}
The interesting contribution comes from the projected Weyl tensor. This contribution from the bulk space, ${\cal E}_{\mu\nu}^{(-1)}$, must be calculated with the 5D Riemann tensor
${^{5}\!R^{(-1)\sigma}_{\mu\tau\nu}}=l_\tau {^{5}\!\dot \Gamma^{(0)\sigma}_{\mu\nu}-l_\nu }{^{5}\!\dot\Gamma^{(0)\sigma}_{\mu\tau}}$.
If we consider $l_\mu l^\mu =0$, i.e., the eikonal equation, then one obtains from Eq.~(\ref{eqn8}) $l^\alpha\bigl(\ddot h_{\alpha\nu}-\frac{1}{2}\bar g_{\alpha\nu}\ddot h\bigr)=0,$
which in other context is used as gauge conditions. It turns out that the contribution from the ${\cal E}_{\mu\nu}^{(-1)}$ don't change these conditions  on $h_{\mu\nu}$.
Let us consider the zero-order  Eq.~(\ref{eqn9}). We need
\begin{eqnarray}
{\cal E}_{\mu\nu}^{(0)}&=&n^\gamma n^\delta {^{4}\!g_\mu^\alpha}{^{4}\!g_\nu^\beta}\Bigl[{^{5}\!R^{(0)}_{\alpha\gamma\beta\delta}}-\frac{1}{3}\Bigl({^{5}\!\bar g_{\alpha\gamma}} {^{5}\!R^{(0)}_{\delta\beta}}-{^{5}\!\bar g_{\alpha\delta}} {^{5}\!R^{(0)}_{\gamma\beta}}
-{^{5}\!\bar g_{\beta\delta}} {^{5}\!R^{(0)}_{\gamma\alpha}}\cr &+&{^{5}\!\bar g_{\beta\delta}} {^{5}\!R^{(0)}_{\gamma\alpha}}\Bigr)
+\frac{1}{12}\Bigl({^{5}\!\bar g_{\alpha\gamma}}{^{5}\!\bar g_{\delta\beta}}-{^{5}\!\bar g_{\alpha\delta}}{^{5}\!\bar g_{\gamma\beta}}\Bigr){^{5}\!R}\Bigr].\label{eqn10}
\end{eqnarray}
Now we take as an example for the radiative coordinates $\Pi(x_\mu)=t+r$ the simplified case $l_\mu =[1,1,0,0,0]$. Then we obtain from the gauge condition that only $h_{11}, h_{22}, h_{44}, h_{55}, h_{13}, h_{14}, h_{15}, h_{34}, h_{45}$ and $h_{35} $ survive.
From the Einstein equations (\ref{eqn9}), one obtains a set of PDE's. When we use the additional gauge conditions: $h_{22}=-h_{11}, h_{34}=h_{35}=h_{45}=h_{14}=h_{15}=0$ (leaving 4
independent $h_{\mu\nu}$ terms), we have 7 unknown functions for the background and first order perturbations: $\bar W_1, \bar\psi, \bar \gamma, \dot h_{13}, \dot h_{11}, \dot h_{44}$ and  $\dot h_{55}$
One  can also  integrate the equation Eq.(\ref{eqn9}) with respect to $\xi$. If we suppose that the perturbations are periodic in $\xi$, we then obtain the Einstein equations with back-reaction terms:
\begin{equation}
{^{4}\!\bar G_{\mu\nu}}=\kappa_4^2{^{4}\!\bar T_{\mu\nu}}+\kappa_5^4 \bar {\cal S}_{\mu\nu}-\bar{\cal E}_{\mu\nu}
+\frac{1}{\tau}\int \Bigl(\kappa_4^2T_{\mu\nu}^{(0)}+\kappa_5^4S_{\mu\nu}^{(0)}-{^{4}\!G_{\mu\nu}^{(0)}}-{\cal E}_{\mu\nu}^{(0)}\Bigr ) d\xi\label{eqn11}
\end{equation}
where we took $\Lambda_{eff}=0$ for the RS fine-tuning and  $\tau$ de period of the high-frequency components.
One can say that the term $-\int {\cal E}_{\mu\nu}^{(0)}d\xi$ in Eq.(\ref{eqn11})is the KK-mode contribution of the perturbative 5D graviton. It is an extra back-reaction term, which contain $\dot h_{55}$ amplified by the warp factor and with opposite sign with repect to the $\kappa_4^2$-term. So it can play the role of an effective cosmological constant.
By substituting back these equations into the original equations, one  gets propagations equations for the first order perturbations. In this way we obtain the set PDE's
\begin{eqnarray}
\partial^2_{tt}\bar W_1=-\partial^2_{rr}\bar W_1+\frac{2}{\bar W_1}(\partial_t\bar W_1^2+\partial_r\bar W_1^2)-\bar W_1(\partial_t\bar \psi^2+\partial_r\bar \psi^2)
+\frac{\bar W_1}{r}(\partial_r\bar \gamma-\partial_t\bar\gamma)\cr
+2(\partial_r\bar W_1-\partial_t\bar W_1)(\partial_t\bar\psi-\partial_r\bar\psi+\partial_r\bar\gamma-\partial_t\bar\gamma)
+2\bar W_1\partial_r\bar\psi\partial_t\bar\psi-4\frac{\partial_r\bar W_1\partial_t\bar W_1}{\bar W_1}
+ 2\partial_{rt}\bar W_1\cr
-\frac{3}{4}\kappa_4^2\Bigl(e^{2\bar\psi}\frac{(\partial_t\bar P-\partial_r\bar P)^2}{\bar W_1 r^2 e^2}
+\bar W_1(\partial_t\bar X-\partial_r\bar X)^2\Bigr),\qquad\qquad\label{eqn12}
\end{eqnarray}
 \begin{eqnarray}
\partial_{tt}\bar\psi=\partial_{rr}\bar\psi +\frac{\partial_r\bar\psi}{r}+\frac{2}{\bar W_1}(\partial_r \bar W_1\partial_r\bar\psi -\partial_t\bar W_1\partial_t\bar\psi)-\frac{\partial_r\bar W_1}{r\bar W_1}\cr
+\frac{3 e^{2\bar\psi}}{4\bar W_1^2r^2e^2}\kappa_4^2\Bigr(\partial_t\bar P^2-\partial_r\bar P^2-\bar W_1^2 e^2\bar X^2\bar P^2 e^{2\bar\gamma-2\bar\psi}\Bigr),\label{eqn13}
\end{eqnarray}
\begin{eqnarray}
\partial_{t}\bar\gamma =\partial_r\bar\gamma+\frac{1}{\partial_t\bar W_1-\partial_r\bar W_1-\frac{\bar W_1}{2r}}\Bigr\{\tfrac{1}{2}\bar W_1(\partial_t\bar\psi -\partial_r\bar\psi)^2+\frac{\partial_r\bar W_1}{r}
-\partial_{tr}\bar W_1+\partial_{rr}\bar W_1 \cr+\frac{2\partial_r\bar W_1\partial_t\bar W_1}{\bar W_1}+(\partial_r\bar W_1-\partial_t\bar W_1)(\partial_r\bar \psi-\partial_t\bar \psi)
-\frac{\partial_r\bar W_1^2+3\partial_t\bar W_1^2}{2\bar W_1}+\kappa_4^2\frac{\bar W_1}{16}\Bigl(7\partial_t\bar X^2
+5\partial_r\bar X^2\cr-12\partial_r\bar X\partial_t\bar X +5\frac{e^{2\bar\gamma}\bar X^2\bar P^2}{r^2}
+6e^{2\bar\psi}\frac{(\partial_r\bar P-\partial_t\bar P)^2}{\bar W_1^2r^2e^2}+\bar W_1^2\beta e^{2\bar\gamma-2\bar\psi}(\bar X^2-\eta^2)^2\Bigr)\Bigr\},\qquad\label{eqn14}
\end{eqnarray}
\begin{equation}
\partial_t\dot h_{13}=\partial_r\dot h_{13}+\ddot k_{13}-\ddot k_{23}+2\Bigl(\frac{\partial_t\bar W_1 -\partial_r\bar W_1}{\bar W_1}+\partial_t\bar\psi -\partial_r\bar\psi\Bigr)\dot h_{13},\label{eqn15}
\end{equation}
\begin{eqnarray}
\partial_t\dot h_{11}=\partial_r\dot h{11}+\frac{e^{2\bar\gamma}}{r^2}\Bigl(\partial_r\bar\psi-\partial_t\bar\psi-\frac{1}{2r}\Bigr)\dot h_{44}+\frac{1}{2}(\ddot k_{22}+\ddot k_{11})-\ddot k_{12}\cr
\frac{2}{\bar W_1}\Bigl(\partial_t\bar W_1-\partial_r\bar W_1+\bar W_1(\partial_r\bar\psi-\partial_t\bar\psi+\partial_t\bar\gamma-\partial_r\bar\gamma)\Bigr)\dot h_{11}\cr
+\frac{1}{2}e^{2\bar\gamma-2\bar\psi}\bar W_1^2\Bigr(\frac{1}{2r}+\frac{\partial_r\bar W_1-\partial_t\bar W_1}{\bar W_1}\Bigr)\dot h_{55}
+\kappa_4^2e^{2\bar\gamma-2\bar\psi}\bar W_1^2(\partial_t\bar X-\partial_r\bar X) \dot\Psi\cos\varphi, \qquad\qquad\label{eqn16}
\end{eqnarray}
\begin{eqnarray}
\partial_\varphi\Bigl(\dot h_{11}+\frac{e^{2\bar\gamma}}{r^2}\dot h_{44}-\bar W_1^2 e^{2\bar\gamma-2\bar\psi}\dot h_{55}\Bigr)+\ddot k_{24}-\ddot k_{14}=
-2\kappa_4^2\bar X\bar Pe^{2\bar\gamma-2\bar\psi}\bar W_1^2\sin\varphi\dot\Psi,\label{eqn17}
\end{eqnarray}
\begin{eqnarray}
\partial_t\dot h_{44}=\partial_r\dot h_{44}+\Bigl(2\partial_r\bar\psi-2\partial_t\bar\psi-\frac{3}{2r}+\frac{\partial_t\bar W_1-\partial_r\bar W_1}{\bar W_1}\Bigr)\dot h_{44}\cr
+\frac{\kappa_4^2}{\epsilon}(\partial_r\bar P-\partial_t\bar P)\dot B+\frac{1}{2}\bar W_1^2 r^2e^{-2\bar\psi}\Bigl(\partial_t\bar\psi-\partial_r\bar\psi+\frac{1}{2r}\Bigr)\dot h_{55}.\label{eqn19}
\end{eqnarray}
We notice that in our simplified case  the equations for the background metric separates from the perturbations. So this example is very suitable to investigate the perturbation equations.
For the first order gauge field perturbation $B_\mu$ we used the condition $l^\mu B_\mu =0$, which is a consequence, as we will see,  of the gauge field equations. So $B_\mu$ can be parameterized as $B_\mu =[B_0,B_0,0,B,0]$.
The propagation equation for $\dot h_{55}$ yields $\dot h_{55}=F_1(t+r)F_2(\varphi,y,\xi)$, which is expected, because the brane part of $\dot h_{55}$ must be separable from the bulk part.
We omitted for the time being, the $\kappa_5^4$ contribution.
It is manifest that to zero order there is an interaction between the high-frequency perturbations from the bulk, the matter fields on the brane and the evolution of $\dot h_{ij}$, also found in the numerical solution{\cite{Slag2:2014,Slag3:2015}. We observe again that the bulk contribution $\dot h_{55}$ is amplified with $\bar W_1^2$. It is a reflection of the massive KK modes felt on the brane.
The equations for the matter fields can be obtained in a similar way. The equation for the background $\bar \Phi$ becomes
\begin{equation}
\bar D^\alpha\bar D_\alpha\bar\Phi-\frac{1}{2}\beta\bar\Phi(\bar\Phi\bar\Phi^*-\eta^2)=\frac{1}{\tau}\int\Bigl(h^{\mu\nu}l_\mu l_\nu\ddot\Psi +\bar g^{\mu\nu}\Gamma^{\alpha(0)}_{\mu\nu}\dot\Psi\Bigr)d\xi.\label{eqn20}
\end{equation}
The equation for $\bar A_\mu$ is the same as in the unperturbed situation. For the first order perturbations we obtain ( for $l^\alpha C_\alpha =0$)
\begin{equation}
\partial_t \dot\Psi=\partial_r\dot\Psi+\frac{\dot\Psi}{\bar W_1}(\partial_r\bar W_1 -\partial_t\bar W_1)+\frac{1}{2r}\dot\Psi,\label{eqn21}
\end{equation}
\begin{eqnarray}
\partial_t\dot B=\partial_r\dot B+\Bigl(\partial_r\bar\psi-\partial_t\bar\psi-\frac{1}{2r}\Bigr)\dot B+
\frac{e^{2\bar\psi}(\partial_t\bar P-\partial_r\bar P)}{2r^2\bar W_1^2 e}\dot h_{44},\label{eqn22}
\end{eqnarray}
\begin{equation}
\partial_t \dot B_0 =\partial_r \dot B_0-\frac{e^{2\bar\gamma}}{r^2}\partial_\varphi \dot B -ee^{2\bar\gamma-2\bar\psi}\dot\Psi \bar X\bar W_1^2\sin\varphi.\label{eqn23}
\end{equation}
For these matter field equations one needs the condition $l^\alpha \bar A_\alpha =0$, otherwise the real and imaginary parts of $\dot \Psi$ interact as the propagation progresses.
From Eq.(\ref{eqn16}), Eq.(\ref{eqn17}) and Eq.(\ref{eqn23}) we observe  on the right hand side  $\varphi$-dependent terms, amplified by $W_1^2$. So the approximate wave solution is no longer axially symmetric, also found by Choquet-Bruhat\cite{Choc:1968}. After integration with respect to $\varphi$, we obtain from Eq.(\ref{eqn17}) ( for $k_{14}=k_{24}$)
\begin{equation}
\dot h_{11}=e^{2\bar\gamma -2\bar\psi}\bar W_1^2\dot h_{55}-\frac{e^{2\bar\gamma}}{r^2}\dot h_{44}-2\kappa_4^2e^{2\bar\gamma-2\bar\psi}\bar X\bar P\bar W_1^2\int(\dot\Psi\sin\varphi) d\varphi.\label{eqn24}
\end{equation}
We notice that the first order disturbance $\dot h_{11}$ (and so $\dot h_{22}=-\dot h_{11}$)) could have its maximum for fixed angle $\varphi$ amplified by the warp factor $W_1^2$.
If we choose for example
$\Psi=\cos(\varphi) \tilde\Psi(t,r,\xi)$, then the last term in Eq.(\ref{eqn24}) becomes $\kappa_4^2\bar X\bar P e^{2\bar\gamma-2\bar\psi}\bar W_1^2\cos(2\varphi )\dot{\tilde\Psi}$, which has  two extremes  on $[0,\pi]$ $\mod(\frac{1}{2}\pi )$. This could be an explanation of the recently found spooky alignment of the rotation axes of quasars over large distances in two perpendicular directions.
\section{Conclusions}
A nonlinear approximation of the field equations of the coupled Einstein-scalar-gauge field equations on a warped 5D spacetime is investigated.  To zeroth order in the expansion parameter it is found that the evolution
of the perturbations on the brane is triggered by the electric part of the 5D Weyl tensor and carries information of the gravitational field outside the brane. The warpfactor in the nominator in front of the bulk contributions will cause a huge disturbance on the brane and could act as dark energy.

\end{document}